\documentclass[preprint,10pt]{article}
\usepackage{amsmath}
\usepackage{enumerate}
\usepackage{array}
\usepackage{graphicx}
\usepackage{float}
\usepackage[footnotesize,bf]{caption}
\usepackage{subcaption}
\usepackage{amsfonts}
\usepackage{epstopdf}
\usepackage[cm]{fullpage}
\usepackage{multicol}
\usepackage[T1]{fontenc}
\usepackage[utf8]{inputenc}
\usepackage{authblk}

\title{Local rewiring rules for evolving complex networks}

\author{E.R. Colman\thanks{E.Colman@Reading.ac.uk}}
\author{G.J. Rodgers}
\affil{Department of Mathematical Sciences, Brunel University, Uxbridge, Middlesex UB8 3PH, U.K.}

\begin{document}
\maketitle
\vspace{-0.7cm}
\begin{abstract}
The effects of link rewiring are considered for the class of directed networks where each node has the same fixed out-degree. We model a network generated by three mechanisms that are present in various networked systems; growth, global rewiring and local rewiring. During a rewiring phase a node is randomly selected, one of its out-going edges is detached from its destination then re-attached to the network in one of two possible ways; either globally to a randomly selected node, or locally to a descendant of a descendant of the originally selected node. Although the probability of attachment to a node increases with its connectivity, the probability of detachment also increases, the result is an exponential degree distribution with a small number of outlying nodes that have extremely large degree. We explain these outliers by identifying the circumstances for which a set of nodes can grow to very high degree.
\end{abstract}

\begin{multicols}{2}
\section{Introduction}
\label{intro}
The question of how complex patterns can be produced by the collective behaviour of many interacting agents such as particles, cells or people, is one of the most important considerations in complexity science. The techniques of statistical physics that originated from the study of gasses and magnets have been adapted to address this question to explain a much wider range of emergent phenomena seen in biological and social systems. Fundamentally, mathematical models are used to derive statistical information about the system as a whole from the assumptions made about its constituent agents, or more specifically, the ``rules'' that govern their interactions. While in most physical systems agents interact with their closest neighbours in a spatial sense, many other systems are not constrained in this way, these are typically modelled as networks where the concept of distance between two points is redefined as the path-length between two nodes. An example of a local rule is triadic closure, the creation of a link between two nodes separated by a path-length of $2$.
\newline

When the growth and evolution of a network is driven by local rules, nodes tend to be selected with a frequency proportional to how well connected they are. This is simply because a node with $x$ connections is present in the neighbourhood of $x$ other nodes, in other words there are $x$ possible ways to discover the node via a local search. It is not suprising then, that the scale-free networks generated by global preferential attachment can also be created by numerous processes that use only local rules i.e. with no global knowledge of the network structure \cite{vazquez}.
\newline

Typically in these models, a network will begin as a small set of nodes connected by edges, then with each iteration, more nodes are introduced and connections made, thus increasing the degree of those that are already there. Networks of this type are partly static in the sense that once an edge has been placed between two nodes it remains in that position for the rest of the network's lifetime. The class of network whose edges are dynamic, i.e. at any point could potentially be removed or rewired, has far wider scope of application.
\newline

This paper studies networks that combine dynamic edges with locally driven processes. Our model is an iterative process that evolves a network, the parameters are the rate of growth, and the rates of local and global (random) rewiring. We examine only networks with directed edges and nodes of a fixed out-going degree. For particular regions of the parameter space, we examine in detail a phenomenon whereby a small set of nodes, owing to their position in the network, gather significantly larger number of connections than those outside the set. These considerations lead to a good approximation of the extreme tail of the degree distribution, giving probabilities for the existence of outlying nodes of the distribution, sometimes refered to as dragon kings \cite{sornette2009dragon}.
\newline

In Section \ref{model} we introduce a model of growth and rewiring in directed networks and show the main results. The following sections describe the mathematical models and their solutions. In Section \ref{random} we find the distribution of cycles of size $n$ in the initial randomly wired graph. In Section \ref{growth} we find a formula for the degree distribution in the large $t$ limit. In Section \ref{dominant} we model the total degree of the dominant nodes and for selected parameter values derive the degree distribution tail.

\section{Related work}
Local rules for growing networks have been in the literature for some time \cite{vazquez, organisation}. 
In the model most similar to the one presented here \cite{PhysRevLett.85.5234}, the preferential attachment mechanism is generalised to include rewiring events. They find both exponential and power-law degree distributions depending on the choice of parameters. Preferential attachment in rewiring has been studied on a network of fixed size with the interesting conclusion that a power law degree distribution can be achieved without a growing network \cite{xie2008scale}. This result relies on the use of a non-linear attachment kernel (heavily biased towards nodes with large degree) to ensure that nodes with large degree continue to grow in spite of the preferential detachment that also occurs through rewiring. This work has been extended to bipartite networks \cite{PhysRevE.72.036120} which have an advantage of being free of degree correlations between neighbouring nodes, thus the results in \cite{evans} for the mean field solution to the degree distribution are exact. The same model also exhibits a condensation phenomenon, also know as gelation \cite{organisation}, where one node becomes connected to almost every other, this is relevant to the study of the dominant nodes presented here.  
\newline

A large body of literature, much of which is commercially motivated, comes from the analysis of the network properties of web $2.0$ systems \cite{ahn2007analysis, mislove2007measurement}. We believe our results here are relevant in this field since rewiring, local dynamics and directed links are present in many of these self-organising systems. Twitter, for example, gives its users the option to ``unfollow'' other users meaning the edges are not static as they are in the majority of complex network models. Local rules, specifically triadic closure contribute to the growth of the network \cite{romero2010directed}, however the distribution does not follow a power-law \cite{kwak2010twitter}.
\newline

Recommendation algorithms designed to facilitate sharing online news articles, music, films etc. connect users together based on the similarity of the content they have responded to positively. The content a user is exposed to in this way is limited to a small number of items shared by her neighbours. When the algorithm updates the links based on the most recent data, we can expect the strength of the similarity between her and her second neighbours to increase, making triadic closure likely. The network topologies of these networks has been studied in \cite{cano2006topology}. In this work the network is treated as a  static object at one instant in time, clustering is found to be significantly higher than the random network which suggests that triadic closure could be part of the networks dynamics. The evolution of a theoretical model network \cite{leadership} considers directed edges between ``leaders'' and ``followers'' that are rewired periodically according to a similarity score. A scale-free structure is found but the authors do not go into detail about the rewiring dynamics. The network evolution of recommendation networks perhaps deserves more attention since it exhibits cumulative advantage effects that have consequences for many commercial areas. 
\newline
 
 Our decision to restrict the model only to the case where every node has the same number of out-going  links was motivated mostly by the considerable simplicity this would bring to the analysis. There is, however, some justification for this assumption regarding the suggested applications. Some product websites link each product to a fixed number of recommended products (amazon.com would be the most famous example although technically the number of recommended products is not fixed as it varies according to the size of the web browser). In the case of Twitter, it is sensible to assume that the number of accounts that a user will follow will, after enough time has passed, remain close to a steady value and not increase to infinity. Each user will differ in the number of other accounts they follow, but if we treat every user as an identical agent with the mean number of followings, then the model we present is appropriate.
 \begin{figure}[H]
  \centering
 \includegraphics[clip,trim=6cm 0cm 0cm 7.5cm, width=0.5\textwidth]{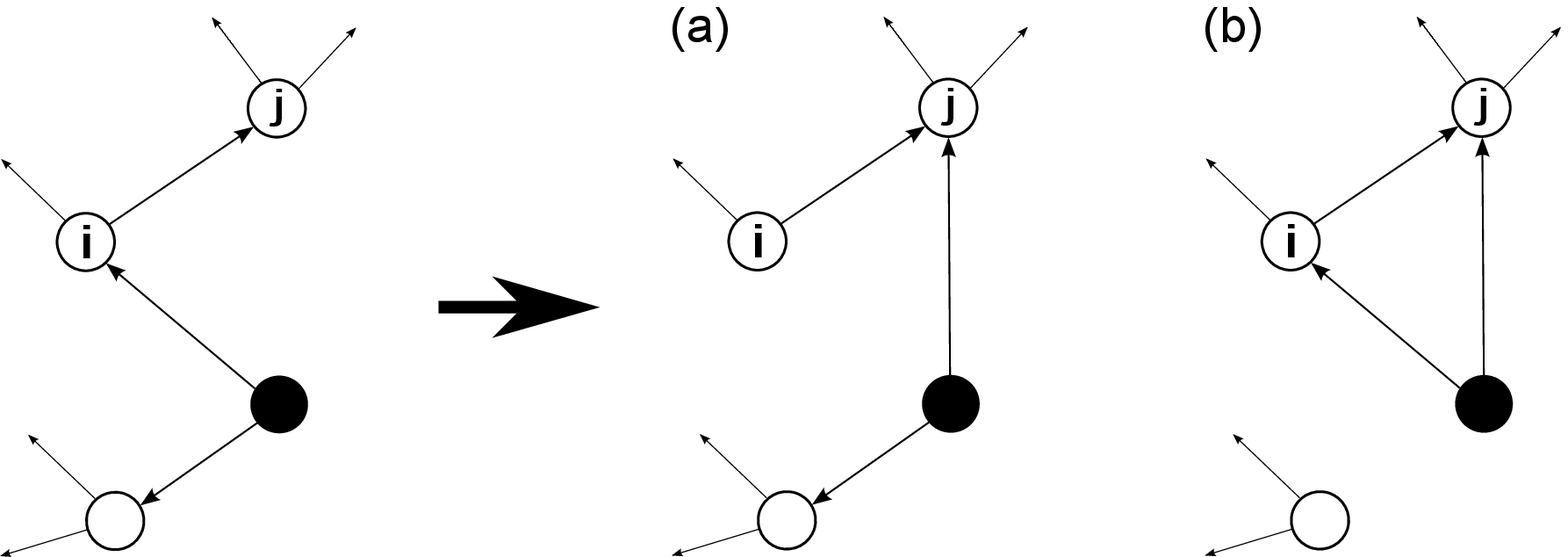}
  \caption{The two possible ways to locally rewire. The left image shows part of the network before rewiring. We consider two possible interpretations of our model. In both, we initially select a random node, in the diagram it is represented by the black node. We then randomly select a target node from all of the nodes that are are a distance of $2$ away from the initially selected node (following the direction of the edges), such as $j$ in the diagram. One of the out-going edges from the black node is then rewired to the target node, it can either be the node that connects the black node to $j$, shown in (a), or it can one which completes the triad, shown in (b).}
  \label{fig:figure1}
\end{figure}
\section{Model and results}
\label{model}
\begin{figure}[H]
 \centering
        \begin{subfigure}[b]{0.37\textwidth}
                \centering
                \includegraphics[width=\textwidth]{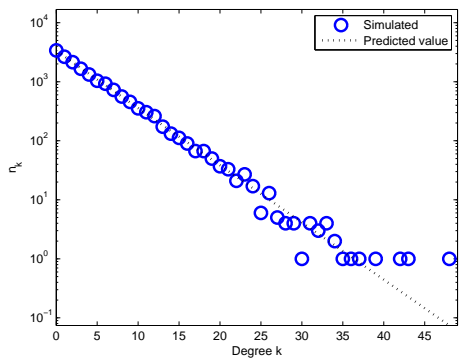}
                \caption{$p=2/3$, $q=1/6$, $m=4$.}
                \label{realSub}
        \end{subfigure}
        \qquad
        \begin{subfigure}[b]{0.37\textwidth}
                \centering
                \includegraphics[width=\textwidth]{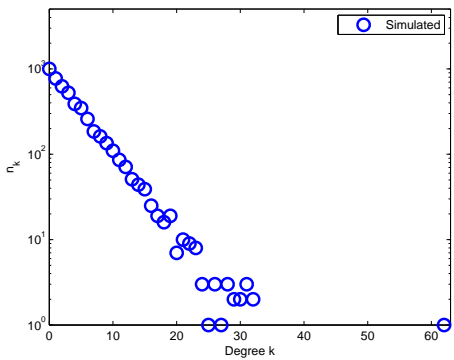}
                \caption{$p=9/10$, $q=1/20$, $m=4$.}
                \label{modelSub}
        \end{subfigure}
\caption{The degree distribution of the network after $10^5$ iterations, starting from an initial random network of $10$ nodes. The line in (\ref{realSub}) shows the predicted result in Eq.(\ref{result1}). In (\ref{modelSub}) an outlier exists owing to the high rate of local rewiring compared with the other mechanisms.} 
\label{fig:figure0}
\end{figure}

\begin{figure}[H]
  \centering
 \includegraphics[width=0.4\textwidth]{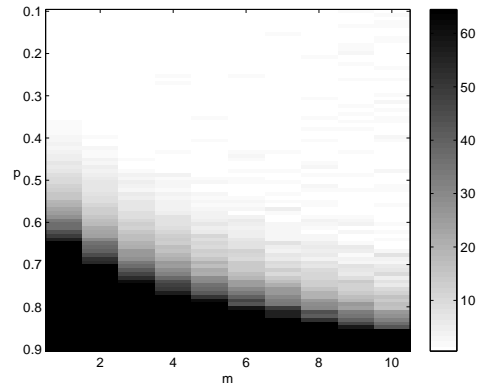}
  \caption{The level of agreement quantified by the Kolmogorov-Smirnov statistic between the prediction for the degree distribution Eq.(\ref{fixed}) and the corresponding numerical simulation. The results presented are for the special case where growth is excluded i.e. when $r=0$ and $q=1-p$, $N=10^{3}$. We consider the model to be accurate up to a KS value of $20$ since this is the value found when we test the prediction of Eq.(\ref{fixed}) against data generated by a pseudo-random numbers drawn from the same probability distribution.}
  \label{heatmap}
\end{figure}

\begin{figure}[H]
 \centering
        \begin{subfigure}[b]{0.46\textwidth}
                \centering
                \includegraphics[width=\textwidth]{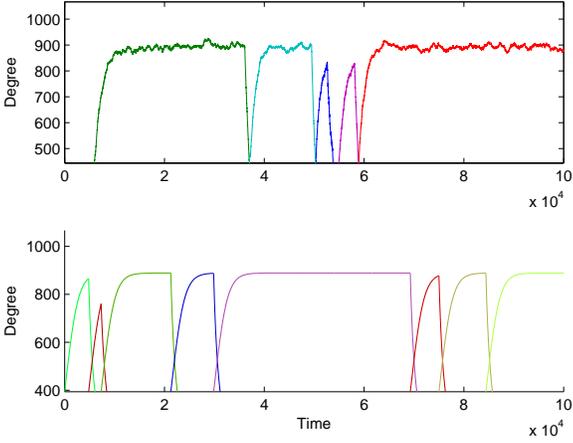}
                \caption{Time dependent dynamics of dominant nodes ($p=0.9$, $N=10^{3}$).}
                \label{time}
        \end{subfigure}
        \qquad
        \begin{subfigure}[b]{0.46\textwidth}
                \centering
                \includegraphics[width=\textwidth]{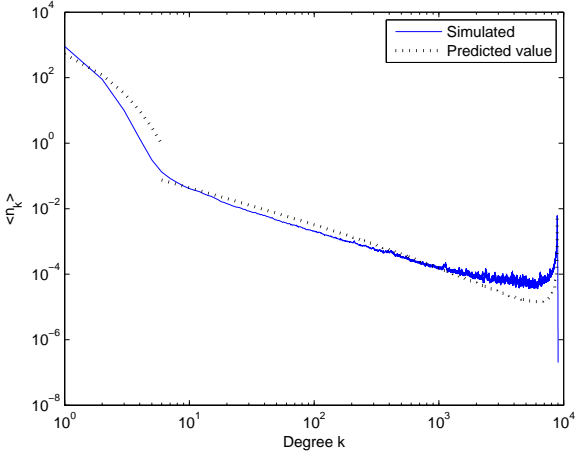}
                \caption{Mean degree distribution of the whole network ($p=0.9$, $N=10^{5}$).}
                \label{tail}
        \end{subfigure}
\caption{Shown here are results when $r=0$ (no growth) and $m=1$. Shown in (\ref{time}) is an example of how the degree of the most connected nodes changes over time (top), and the equivalent approximation using the method outlined in \ref{time_dyn}. Each colour represents a different node. The distribution is divided into two regimes; the exponential part when $\langle n_{k} \rangle\geq 1$ and the tail. The prediction comes from Eq.(\ref{adjusted}) for the first part and Eq.(\ref{shape}) for the second.}
\end{figure}

Let $G(N,mN)$ be a random graph in which each of the $N$ nodes has $m$ out-going directed edges, the destination of each directed edges is selected randomly. Throughout this paper we use `degree' to refer to the in-coming degree of a node. In each time-step the network develops in one of the following ways
\begin{itemize}
\item \textbf{Local rewiring:} With probability $p$, randomly select a node and rewire one of its out-going edges to a randomly selected descendant of one of its descendants (see Fig.(\ref{fig:figure1})).

\item \textbf{Global rewiring:} With probability $q$,  randomly select a node and rewire one of its out-going edges to a randomly selected node.

\item \textbf{Growth:} With probability $r$, introduce a node to the network with $m$ out-going edges, attach the edges to randomly selected nodes in the network.
\end{itemize}
For convenience we set $r=1-p-q$. As we iterate this process, the binomial degree distribution of the initial network converges towards an exponential distribution for every choice of $p,q$ and $m$ (Fig.(\ref{fig:figure0})). When $q$ is small and $p$ is relatively large we observe additional dynamics where we see a small number of outlying nodes with degrees much higher than predicted by the exponential distribution (Figures (\ref{modelSub}) and (\ref{heatmap})). These are the conditions for ``rich-clubs'' to develop, small sets of nodes whose growth in degree is magnified by the fact that the set has very few out-going links. 
The outlying nodes, which we call `dominant nodes', exist because their out-going edges belong to small cycles. This is illustrated most easily in the case where $m=1$; over time the outliers increase in degree until the cycle they belong to is broken, at this point the degree rapidly falls while a new dominant node begins its rise (Fig.(\ref{time})). For sufficiently small $q$, the node remains dominant long enough to reach a state where its degree, on average, is neither increasing or decreasing, this causes a small spike in the tail of the degree distribution (Fig(\ref{tail})).

\section{Random graphs with directed edges and fixed out-degree}
\label{random}
For the $mN$ edges in the network, each is attached to the node $i$ with probability $1/N$ the probability that $i$ has degree $k$ is the probability of $k$ successes in $mN$ trials. Letting $P_{k}$ denote the probability that any node has degree $k$ we have
\begin{equation}
P_{k}=\binom{mN}{k}\left(\frac{1}{mN}\right)^{k}\left(1-\frac{1}{mN}\right)^{N-k}.
\end{equation}
Let $l_{i,j}$ be the length of a path from node $i$ to node $j$ where no nodes are visited more than once, and let $L_{n}$ be the average number of such paths that have $l_{i,j}=n$.
 We can find solutions for the average of $L_{n}$ over the network ensemble from the recursion
\begin{equation}
L_{n}=L_{n-1}\frac{m(N-n)}{N}.
\end{equation}
The fraction on the right hand side is the probability that the next edge in the path does not link to any of its ancestor nodes in the path or to itself. We have $L_{0}=N$ so
\begin{equation}
L_{n}=\frac{m^{n}}{N^{n-1}}\frac{(N-1)!}{[N-(n+1)]!}.
\end{equation}
This also gives a formula for the average number of cycles $C_{n}$ of length $n$
\begin{equation}
nC_{n}=L_{n-1}\frac{m}{N}
\end{equation}
giving
\begin{equation}
C_{n}=\frac{m^{n}}{nN^{n-1}}\frac{(N-1)!}{(N-n)!}.
\end{equation}
It is important to note that every network in this class will have at least one cycle and that every node either belongs to a cycle or is connected to a cycle by a directed path. 
\section{Degree distribution}
\label{growth}
In a single time-step the probability of attaching to a node $i$ with degree $k_{i}$ is
\begin{equation}
\label{attachment}
\Pi_{a}(k_{i})=p\left[\frac{k_{i}}{mN}\right]+q\left[\frac{1}{N}\right]+r\left[\frac{m}{N}\right].
\end{equation}
This assumes that node degree correlations do not effect the attachment probability, i.e. the degree of a parent node of $i$ is approximated well by the mean degree $m$. Therefore the number of edges that can potentially be rewired to $i$ is $mk_{i}$, multiplying by the probability $1/m$ that once selected, $i$ will be the node redirected to gives the first term on the left hand side of Eq.(\ref{attachment}). The probability of removing an adjacent edge from $i$ is
\begin{equation}
\Pi_{r}(k_{i})=(p+q)\left[\frac{k_{i}}{mN}\right].
\end{equation}
We are interested in finding $n_{k}(t)$, the number of nodes with in-coming degree $k$. At $k=0$
\begin{equation}
\label{rate0}
\frac{\partial n_{0}}{\partial t}=r+\frac{(p+q)}{mN}n_{1}-\frac{q+rm}{N}n_{0}.
\end{equation}
The terms on the right hand side respectively represent the addition of a node to the network, creation of a node of degree $0$ by removing an edge from a node of degree $1$, and destruction by attaching an edge and making it a node of degree $1$. Similarly for $k\geq 1$,
\begin{equation}
\label{rate1}
\begin{split}
\frac{\partial n_{k}}{\partial t}=&\frac{(p+q)}{mN}[(k+1)n_{k+1}-kn_{k}]\\
&+\frac{p}{mN}[(k-1)n_{k-1}-kn_{k}]\\
&+\frac{q+rm}{N}[n_{k-1}-n_{k}].
\end{split}
\end{equation}
The first pair of terms on the right hand side represent the mean change in $n_{k}$ by either creating or destroying a node of degree $k$ by removing one of its edges, the second pair are similar except for attachment by local rewiring, the third is for global rewiring.

As $t$ grows large, the proportion of node of degree $k$ will converge to constant values. Therefore in the asymptotic limit as $t\rightarrow \infty$ Eq.(\ref{rate1}) reduces to the following second order recursion relation, found by substituting $N(t)=rt$ and $P_{k}=n_{k}(t)/N$.
\begin{equation}
\label{rec}
\begin{split}
\left[r+(q+rm)+\frac{2p+q}{m}k\right]P_{k}=[&(q+rm)+\frac{p}{m}(k-1)]P_{k-1}\\
&+\frac{p+q}{m}(k+1)P_{k+1}
\end{split}
\end{equation}
and Eq.(\ref{rate0}) becomes
\begin{equation}
\label{zero}
\left[q+(m+1)r\right]P_{0}=r+\frac{p+q}{m}P_{1}.
\end{equation}
We introduce the generating function
\begin{equation}
\label{gen}
g(x)=\sum_{k=0}P_{k}x^{k},
\end{equation}
following the method outlined in Appendix \ref{generator} we get
\begin{equation}
\begin{split}
\left[\frac{-p}{m}x^{2}+\frac{2p+q}{m}x-\frac{p+q}{m}\right]&g'(x)\\
+[-(q+rm)x+&r+q+rm]g(x)\\
=&-\frac{p+q}{m}P_{1}+(r+q+rm)P_{0}.
\end{split}
\end{equation}
The right hand side equates with Eq.(\ref{zero}) to give
\begin{equation}
\label{diff}
\frac{p}{m}(1-x)\left(x-\frac{p+q}{p}\right)g'(x)+[r+(q+rm)(1-x)]g(x)=r
\end{equation}
for $q,r \neq 0$. The solution is

\begin{equation}
\label{withmu}
\begin{split}
g(x)=&\\
-\frac{rm}{p}&(x-(p+q)/p)^{-\mu}\sum_{n=0}^{\infty}\binom{\mu-1}{n}\left(-\frac{q}{p}\right)^{\mu-n-1}\frac{(x-1)^{n}}{n+\lambda}
\end{split}
\end{equation}
where
\begin{equation}
\lambda(p,q,m)=\frac{rm}{q}\\
\end{equation}
and
\begin{equation}
\mu(p,q,m)=m\left[\frac{q+rm}{p}-\frac{r}{q}\right].
\end{equation}
Notice that the terms in $\lambda$ and $\mu$ are simply the ratios of the different rates of attachment by the three different mechanisms in the process.   
To return the degree distribution $P_{k}$ we equate the coefficients of $x^{k}$ in the expansion of $g(x)$with Eq.(\ref{gen}). This is easily done when $\mu$ is a positive integer, for example when $\mu=1$,
\begin{equation}
\label{result1}
P_{k}=\frac{q}{p+q}\left(\frac{p}{p+q}\right)^{k}
 \end{equation}
and $\mu=2$
\begin{equation}
\begin{split}
P_{k}=&\\
\frac{m(1-p-q)}{(p+q)^{2}}&\left(\frac{p}{p+q}\right)^{k}\left[\left(\frac{q}{1+\lambda}-\frac{q}{\lambda}\right)k-\left(\frac{p}{1+\lambda}+\frac{q}{\lambda}\right)\right].
\end{split}
\end{equation}
In fact when $\mu$ is any positive integer the form of $P_{k}$ is the product of an exponential part and a polynomial in $k$ of order $\mu$. In the case of a  network with fixed size $N$, $r=0$, we solve Eq.(\ref{diff}) to find
\begin{eqnarray}
\label{fixed}
P_{k}=\frac{p^{k}(1-p)^{m(1-p)/p}}{k!}\left(k-1-\alpha\right)\left(k-2-\alpha\right)...\left(-\alpha\right) \nonumber
\end{eqnarray}
where
\begin{equation}
\alpha=-\frac{(1-p)m}{p}.
\end{equation}
An interesting result occurs when we set the parameter values in terms of $m$, 
\begin{equation}
p=\frac{m}{m+2}\text{ and }q=r=\frac{1}{m+2}.
\end{equation}
The generating function in this case is found to be
\begin{equation}
g(x,m)=\frac{1}{(m+1)-mx}
\end{equation}
which gives the result
\begin{equation}
P_{k}=\frac{1}{m+1}\left(\frac{m}{m+1}\right)^{k}.
\end{equation}
Remarkably, this is exactly the result found in \cite{bollobas2001degree} for the uniform attachment model, which is also a specialisation of the present model when $p=q=0$. 

\section{Dominant nodes}
\label{dominant}
Consider the extreme example where $m=1$ and $p=1$, the steady state solution for the degree distribution is a network comprising of one node of degree $N$ which is linked to by every node the network including itself. Hence, as $p$ approaches $1$ we anticipate the existence of nodes with degree much higher than predicted in Section \ref{growth}, and a possible alteration to the topology of the entire network. The mathematical formulation of the model in Section \ref{growth} (Equations (\ref{rate0}) and (\ref{rate1})), did not account for this and so we model specifically the degree of the nodes which are likely to dominate the network. Previous work has examined the similar concept of gelation, where a gel node takes a finite proportion of the network's $N$ nodes as $N$ goes to infinity \cite{organisation,evans}. To become dominant a node must belong to a subset of nodes called a ``rich-club''; a small set of nodes characterised by the large number of links between its members relative to the small number of links that leave the set \cite{colizza2006detecting}. In this section we present the equation that describes the dynamics of the total degree of the rich-club before taking a detailed look at the simplest case, when $m=1$ and $r=0$.
\newline

Let $R$ be a subset of $n_{R}$ nodes, let $k_{R}^{\text{in}}(t)$ denote the total number of in-coming edges adjacent to $R$ and $k_{R}^{out}(t)$ the number of out-going edges. Using a continuum approximation
\begin{equation}
\label{rich}
\frac{\partial k_{R}^{\text{in}}}{\partial t}=[q+rm]\frac{n_{R}}{N}+p\frac{k_{R}^{\text{in}}}{m^{2}N}\left(\frac{mN-k_{R}^{\text{in}}}{N}\right)-\left(\frac{k_{R}^{\text{out}}}{mn_{R}}p+q\right)\frac{k_{R}^{\text{in}}}{mN}
\end{equation}
The first term on the left hand side comes from attachment during growth or global rewiring, the second term comes from local rewiring and is the product of the probability that a second neighbour of $R$ is selected, and the the probability that once selected it will rewire to $R$ (it assumes only one edge exits from the neighbour to $R$), the last term shows the decrease when one of the edges coming into $R$ is rewired away, $k_{R}^{\text{out}}/mn_{R}$ is the probability that the edge which guides the local rewiring is one that leaves the set $R$. When $N>>n_{R}$ and $O(1/N)$ terms are disregarded Eq.(\ref{rich}) becomes
\begin{equation}
\label{cont}
\frac{\partial k_{R}^{\text{in}}}{\partial t}=\frac{k_{R}^{\text{in}}}{m^{2}N}\left(p-p\frac{k_{R}^{\text{in}}}{N}-qm-p\frac{k_{R}^{\text{out}}}{n_{R}}\right)
\end{equation}
 If a set $R$ exists such that this derivative is positive, i.e. if
\begin{equation}
k_{R}^{\text{in}}>N\left(\frac{qm}{p}-\left[1-\frac{k_{R}^{\text{out}}}{n_{R}}\right]\right)
\end{equation}
then the nodes in $R$ will begin to dominate the network. However, the edges in this model are transient, and $R$ will only maintain its structure until one of its internal edges is selected for rewiring. 
\subsection{Rich-club structure}
Rich-clubs are characterised by a large number of internal links relative to their number of nodes. It is therefore likely that such sets will contain small cycles. Nodes which have out-going edges that link back on themselves have less chance of losing adjacent edges from local rewiring than those which don't, and the same can be said for reciprocated links (cycles of length $2$). We are therefore interested in the dynamics of cycles. Each cycle of length $n$ will exist for precisely $\Delta t$ iterations with probability
\begin{equation}
\pi_{n}(\Delta t)=\left[1-\frac{(p+q)n}{mN}\right]^{\Delta t}\frac{(p+q)n}{mN}
\end{equation}
giving a mean lifespan of
\begin{equation}
\label{mean_cycle}
\langle \Delta t \rangle_{n}=\frac{mN-(p+q)n}{(p+q)n}.
\end{equation}
These formula give some indication of the structure of the network, particularly in those subsets of nodes that are highly interconnected, however, modelling the evolution of a rich-club is an intricate problem. We continue by investigating only the simple case where $r=0$, $m=1$ and $q=1-p$.

\subsection{$r=0$, $m=1$ and $q=1-p$}  
Suppose $R$ is a single node. Let $k(t)=k_{R}^{\text{in}}(t)$. The solution to Eq.(\ref{cont}) is
\begin{equation}
\label{t_down}
T_{\text{down}}(k_{\tau},k)=\frac{N}{1-p}\ln\left[\frac{k_{\tau}}{N(1-p)/p+k_{\tau}}\frac{N(1-p)/p+k}{k}\right].
\end{equation}
Here $T_{\text{down}}(k_{\tau},k)$ represents the average time taken for $R$ to decrease from degree $k_{\tau}$ to $k$. Suppose $R$ is self-cyclic (meaning that its one outgoing edge links back on itself). Now, if an edge adjacent to $R$ is selected for local rewiring it will be rewired to exactly the position it was in initially. The solution to Eq.(\ref{cont}) becomes
\begin{equation}
\label{t_up}
T_{\text{up}}(k_{\tau},k)=\frac{N}{2p-1}\ln\left[\frac{N(2p-1)/p-k_{\tau}}{k_{\tau}}\frac{k}{N(2p-1)/p-k}\right].
\end{equation}
Here $T_{\text{up}}(k_{\tau},k)$ represents the average time taken for $R$ to increase from degree $k_{\tau}$ to $k$.


To predict the tail of the degree distribution $\langle n_{k}\rangle$ we assume that it is proportional to the expectation of the length of time for which a dominant node has degree $k$. Suppose $i$ is a node of degree $k_{i}$ which becomes self-cyclic. The probability that the degree of $i$ will grow to size $k+1$ or greater is the probability that $i$ will not be selected for global rewiring in $T_{\text{up}}(k_{i},k+1)$ consecutive iterations. Given that this occurs, the total time for which $i$ has degree $k$ is given by Eqs.(\ref{t_down}) and (\ref{t_up}). Putting this together we get
\begin{equation}
\label{shape}
\langle n_{k}\rangle \approx C\left(1-\frac{1-p}{N}\right)^{T_{\text{up}}(k_{i},k+1)}[T_{\text{up}}(k,k+1)+T_{\text{down}}(k+1,k)]
\end{equation}
where $C$ is the constant of proportionality and depends on $k_{i}$. In Appendix \ref{time_dyn} we show how the mean of $k_{i}$ can be approximated and the results are plotted in Fig.(\ref{tail}). Eq.(\ref{shape}) only approximates the shape of the tail of the degree distribution, it should be noted that we have neglected the time for which a node has degree $k$ but $\langle n_{k}\rangle$ does not reach degree $k+1$, for this reason $\langle n_{k}\rangle$  quickly approaches infinity as $k$ approaches its upper bound. 

\subsection{Effect on the rest of the network}
Previously we have used the mean degree to approximate the number of second neighbours of any given node, and hence the attachment probability for local rewiring. In cases where a significant proportion of the edges are attached to a small number of dominant nodes the expectation of the number of second neighbours of a node is less and Eq.(\ref{fixed}) fais to give an accurate prediction (see Fig.(\ref{heatmap})). If we let $\langle k_{-e} \rangle$ be the mean degree of the network excluding any number of edges then the equivalent of Eq.(\ref{attachment}) is
\begin{equation}
\Pi_{a}(k_{i})=(1-p)\frac{1}{N}+p\frac{\langle k_{-e} \rangle k_{i}}{m^{2}N}
\end{equation}
which gives
\begin{equation}
\label{steady}
\begin{split}
mN\frac{\partial n_{k}}{\partial t}=&(k+1)n_{k+1}-kn_{k}+(1-p)m[n_{k-1}-n_{k}]\\
+\frac{p\langle k_{-e} \rangle}{m}&[(k-1)n_{k-1}-kn_{k}].
\end{split}
\end{equation}
This can be solved in a similar way to before, but since we are not considering growth we can adopt a simpler method, used in \cite{xie2008scale}, and assume that for large $t$ a steady state has been reached and the left hand side is $0$. Eq.(\ref{steady}) can be rewritten
\begin{equation}
\begin{split}
kn_{k}-(k+1)n_{k+1}=&\left[(1-p)m+\frac{p\langle k_{-e} \rangle}{m}(k-1)\right]n_{k-1}\\
-&\left[(1-p)m+\frac{p\langle k_{-e} \rangle}{m}k\right]n_{k}.
 \end{split}
 \end{equation}
We immediately see that
\begin{equation}
kn_{k}=\left[(1-p)m+\frac{p\langle k_{-e} \rangle}{m}(k-1)\right]n_{k-1}
 \end{equation}
and so we find
\begin{equation}
\label{adjusted}
n_{k}=\left(\frac{p\langle k_{-e} \rangle}{m}\right)^{k}\frac{1}{k!}\left(k-1-\alpha\right)\left(k-2-\alpha\right)...\left(-\alpha\right)n_{0}
\end{equation}
where
\begin{equation}
\alpha=-\frac{(1-p)m^{2}}{p\langle k_{-e} \rangle}.
\end{equation}
Knowing that the sum over all $k$ is $N$ we also find
\begin{equation}
n_{0}=N(1-p)^{-\alpha}.
\end{equation}

\section{Conclusion}
\label{remarks}
The model presented is one of the simplest possible treatments of rewiring in directed networks and although we have not related it to any particular application, these results add to the understanding of this class of network as a whole. We have looked at local rules that naturally lead to the preferential selection of nodes for attachment, and global rules that select nodes randomly. Edges are selected with equal probability for rewiring which leads to nodes being selected proportionally to their degree. The combined effect of the these two mechanisms is a network with predominantly a exponential degree distribution. The vast majority of nodes do not accumulate edges to create a long (power-law) tail. Instead we find a small number of dominant nodes who conspire to develop an immunity to local detachment causing a large number of links to condense around them. 

\section*{Acknowledgements}
ERC is grateful for the financial support of the EPSRC.

\bibliography{bibfile2}
\bibliographystyle{ieeetr}

\appendix
\section{Solving the first order recursion relation}
\label{generator}
For the recursion relation
\begin{equation}
[m_{0}k+c_{0}]P_{k}+[m_{-1}(k-1)+c_{-1}]P_{k-1}+[m_{1}(k+1)+c_{1}]P_{k+1}=0
\end{equation}
first multiply by $x^{k}$
\begin{equation}
\begin{split}
m_{0}kP_{k}x^{k}+c_{0}P_{k}x^{k}+m_{-1}(k-1)P_{k-1}x^{k}+c_{-1}P_{k-1}x^{k}&\\
+m_{1}(k+1)P_{k+1}x^{k}+c_{1}P_{k+1}x^{k}&=0.
\end{split}
\end{equation}
Rewrite this as
\begin{equation}
\begin{split}
xm_{0}kP_{k}x^{k-1}+c_{0}P_{k}x^{k}+x^{2}m_{-1}(k-1)P_{k-1}x^{k-2}&\\
+xc_{-1}P_{k-1}x^{k-1}+m_{1}(k+1)P_{k+1}x^{k}+x^{-1}c_{1}P_{k+1}x^{k+1}&=0.
\end{split}
\end{equation}
Summing over $k \geq 1$
\begin{equation}
\begin{split}
xm_{0}\sum_{k=1}kP_{k}x^{k-1}+c_{0}\sum_{k=1}P_{k}x^{k}+x^{2}m_{-1}\sum_{k=0}kP_{k}x^{k-1}&\\
+xc_{-1}\sum_{k=0}P_{k}x^{k}+m_{1}\sum_{k=2}kP_{k}x^{k-1}+x^{-1}c_{1}\sum_{k=2}P_{k}x^{k}&=0.
\end{split}
\end{equation}
Introduce the generating function
\begin{equation}
g(x)=\sum_{k=0}P_{k}x^{k}
\end{equation}
and we have
\begin{equation}
\begin{split}
xm_{0}g'(x)+c_{0}[g(x)-P_{0}]+x^{2}m_{-1}g'(x)+xc_{-1}g(x)&\\
+m_{1}[g'(x)-P_{1}]+x^{-1}c_{1}[g(x)-P_{0}-P_{1}x]&=0
\end{split}
\end{equation}
or
\begin{equation}
\begin{split}
[x^{2}m_{-1}+xm_{0}+m_{1}]g'(x)+[xc_{-1}+c_{0}+x^{-1}c_{1}]g(x)&\\
=[m_{1}+c_{1}]P_{1}+[c_{0}+c_{1}x^{-1}]P_{0}&
\end{split}
\end{equation}
Since $g(1)=1$ and $g'(1)=\langle k \rangle$
\begin{equation}
[m_{-1}+m_{0}+m_{1}]\langle k \rangle+[c_{-1}+c_{0}+c_{1}]=[m_{1}+c_{1}]P_{1}+[c_{0}+c_{1}]P_{0}
\end{equation}

\section{Estimating the mean degree of dominant nodes}
\label{time_dyn}
We consider a model that describes the time dependent behaviour of the dominant nodes with the following simplifying assumptions:
\begin{enumerate}
\item At any time there will be exactly one self-cyclic node whose degree increases according to Eq.(\ref{t_up}).
\item The times for which nodes remain self-cyclic are geometrically distributed with mean $N/(1-p)$.
\item After the out-edge of a self-cyclic node is rewired globally its degree decreases according to Eq.(\ref{t_down}).
\end{enumerate}
Additionally we assume that the degree of a node when it initially becomes self-cyclic is $k_{0}$, which we find by simultaneously solving
\begin{equation}
\label{top}
k_{\text{top}}=\frac{N\beta}{\left(\dfrac{N\beta}{k_{0}}-1\right)\exp\left(\dfrac{1-2p}{1-p}\right)+1}
\end{equation}
where $k_{\text{top}}$ is the degree of a self-cyclic node after the average amount of time it remains cyclic (from Eq.(\ref{t_up})), $\beta=(2p-1)/p$, and
\begin{equation}
\label{approx}
k_{0} \approx \frac{k_{\text{top}}^{2}}{2N}.
\end{equation}
To understand this approximation consider that when global rewiring of the self-cyclic node occurs, it may rewire to form a $2$-cycle with probability $k_{t}/N$, then when local rewiring happens on one of the edges in the $2$-cycle a self-cyclic node is created and the expectation of its degree is $k_{t}/2$. If this does not occur then we assume that the new self-cyclic node has small degree (close enough to $0$ to be ignored). Eq.(\ref{approx}) is the expected outcome of those two possibilities.
Solving Eqs. (\ref{top}) and (\ref{approx}) gives
\begin{equation}
k_{0}=\frac{N}{2}\left(\frac{\beta}{2(1-\theta)}\left[1+\sqrt{1-\frac{8\theta(1-\theta)}{\beta}}\right]\right)^{2}
\end{equation}
where
\begin{equation}
\theta=\exp\left(\dfrac{1-2p}{1-p}\right).
\end{equation}
Through numerical investigation we determine that $k_{0}$ is real valued for $p>0.77$.

The expectation of the number of nodes that have degree $k>k_{0}$ at any time $t$ is given by the length of time a self-cyclic node has degree $k$ divided by the mean length of time a node remains self-cyclic. For $k>k_{0}$,
\begin{equation}
n_{k}\approx \frac{1-p}{N}\left(1-\frac{1-p}{N}\right)^{T_{\text{up}}(k_{0},k+1)}[T_{\text{up}}(k,k+1)+T_{\text{down}}(k+1,k)].
\end{equation}
The average number of edges linking to dominant nodes is
\begin{equation}
\langle k_{-e} \rangle=\sum_{k'=k_{0}}^{N} k'n_{k'}.
\end{equation}
Fig.(\ref{time}) compares the model described here, and the mean found from simulating the actual model.

\end{multicols}

\end{document}